\documentclass[12pt]{iopart}

\usepackage{amssymb}
\usepackage{graphicx} 
\usepackage{multirow}
\usepackage{fullpage}
\usepackage{tabularx}
\usepackage{dcolumn}
\usepackage{booktabs}
\usepackage{rotating, makecell}
\usepackage{float}
\usepackage{hyperref}
\usepackage{multicol}
\usepackage{subfig}
\usepackage{derivative}

\begin{document}

\title{Epitaxial thin film growth in the U-Ge binary system}

\author{Syed Akbar Hussain, Ali A. M. H. Jasem, Lottie M. Harding, Ross S. Springell, Christopher Bell}

\address{School of Physics, University of Bristol, Tyndall Avenue, Bristol, BS8 1TL, United Kingdom}
\ead{christopher.bell@bristol.ac.uk}

\begin{abstract}
We explore the U-Ge phase diagram using thin film growth by co-deposition of U and Ge via d.c. magnetron sputtering. Using three different single crystal substrates - MgO, CaF$_2$ and SrTiO$_3$ - we have stabilised mixed phase films of UGe$_3$ and UGe, with evidence of UGe$_2$ crystallisation. At higher temperatures UO$_2$ forms as a consequence of gettering of oxygen from several types of substrate. Several UGe$_3$ dominated samples grown on MgO substrates have also been characterised electrically, showing residual resistivity ratios up to six. We discuss future directions for fundamental research for this intriguing material system in thin film form.
\end{abstract}


\section{Introduction}
\label{introduction}
Binary compounds incorporating uranium display a wide range of physics, from metallic ferromagnetism and antiferromagnetism in UH$_3$ \cite{troc_uh3_1995} and UN \cite{curry_un_1965}, respectively, to exotic superconductivity and heavy fermion character in systems such as UPt$_3$ \cite{joynt_superconducting_2002} and UGe$_2$ \cite{saxena_uge2_nature2000}, to name but a few examples. In this context the materials science of their diverse crystal structures is of great interest. In bulk many of these systems have been studied in detail, but far less work has been done on thin film systems \cite{springell_review_2022}. Thin films not only offer the potential to create idealised surfaces for study, but possess a range of tuning parameter that can be used to control the structural, magnetic and electronic properties of materials, such as the dimensionality (via film thickness), substrate strain and proximity with other materials \cite{chatterjee_hf_review_2021}.

Following a recent study of thin film epitaxial growth of U-Si materials \cite{harding_usi_2023}, here we study the U-Ge binary system, motivated by the possibility of stabilizing phase-pure single crystal thin films of UGe$_2$ to ultimately allow the study of its unconventional superconductivity, magnetism, quantum criticality and heavy fermion behavior. To achieve successful epitaxial thin film growth of materials whose binary phase diagrams have a range of line compounds, appropriate single crystal substrates need to be identified. These should provide a suitable lattice match to the desired phase, as well as being chemically compatible - minimizing interdiffusion or chemical reactions between the film and substrate during growth, which is often required to be at elevated temperatures to obtain good crystallinity. Additionally, a key challenge is to find substrates and growth conditions that promote the formation of one phase.  

Before introducing the experimental methods, we briefly discuss the various crystal structures and electronic properties in U-Ge phase diagram, compare with the U-Si system, and show some possible lattice matches with several common commercial single crystal substrates. The U-Ge binary phase diagram consists of five different phases namely: U$_5$Ge$_4$, UGe, U$_3$Ge$_5$, UGe$_2$ and UGe$_3$ in order of increasing Ge content. These are summarised in Table \ref{table_uge}, which also shows the symmetries and lattice parameters of these phases. We note that the original phase diagram by Lyahsenko \& Bykov \cite{lyahsenko_u_ge_1960} included U$_7$Ge and U$_3$Ge$_4$ phases which were later corrected to U$_3$Ge$_5$ and UGe. Compared to the U-Si system, the line compounds are slightly different in stoichiometry. In the U-Si case there are the following phases: U$_3$Si, U$_3$Si$_2$, USi, U$_3$Si$_5$, USi$_2$ and USi$_3$ in order of increasing Si content \cite{middleburgh_jnm_2016}. Given the smaller Si atom compared to Ge, the U-Si bond length is shorter than the U-Ge bond length, leading to a stronger $5f-$ligand hybridisation in the silicides, meaning that the electrons are more itinerant in U-Si phases compared to U-Ge which are more localised. 

As a consequence, several of the U-Ge compounds are magnetic. For example a ferromagnetic transition at $94$ K, is found in U$_3$Ge$_5$. This difference can also be framed in the Hill limit picture \cite{hill_1970}, since the shortest distance U-U bond length, $d_{\mathrm{U-U}}$, in U$_3$Ge$_5$ is $3.95$ \AA ~which is well above the Hill limit of $3.4$ \AA. Following a similar argument $d_{\mathrm{U-U}}\sim 3.84$ \AA ~for UGe$_2$, also places it in the localised electron behaviour region, consistent with the observed ferromagnetic state. In the case of the U-rich germanides, both U$_5$Ge$_4$ and UGe exhibit paramagnetic behaviour down to $2$ K, with $d_{\mathrm{U-U}} \sim 2.9$ \AA ~placing them well below the Hill limit. 
\begin{table*}
 \centering
 \begin{tabular}{|c|c|c|c|c|c|c|} 
 \toprule
       Phase  & Structure & Space group & $a$ (\AA) & $b$ (\AA) & $c$ (\AA) & Reference  \\
       \midrule
        U$_5$Ge$_4$ & Hexagonal & P63/mcm & 8.744 & 8.744 & 5.863 & \cite{boulet_jac_1997}  \\
       UGe  & Orthorhombic & Pbcm  & 9.827 & 8.932 & 5.841 & \cite{troc_magnetotransport_2002}  \\
        U$_3$Ge$_5$ & Hexagonal & P6/mmm  & 3.954 & 3.954 & 4.125 &  \cite{boulet_crystal_1999}  \\
        UGe$_2$ & Orthorhombic & Cmmm & 3.997 & 15.039 & 4.087 & \cite{saxena_uge2_nature2000} \\ 
       UGe$_3$   & Cubic & Pm-3m & 4.206 & 4.206 &  4.206 &  \cite{lander_magnetization_1979}  \\ 
        \bottomrule
  \end{tabular}
 \caption{Different phases present in the U-Ge binary phase diagram, together with crystal symmetries, space groups and lattice parameters ($a$, $b$ and $c$). Phases are listed in order of increasing Ge content, from top to bottom. \label{table_uge}}
\label{tab:chap7/UGe Phases}
\end{table*}

\begin{figure}
\centering
	\includegraphics[width=0.35\textwidth]{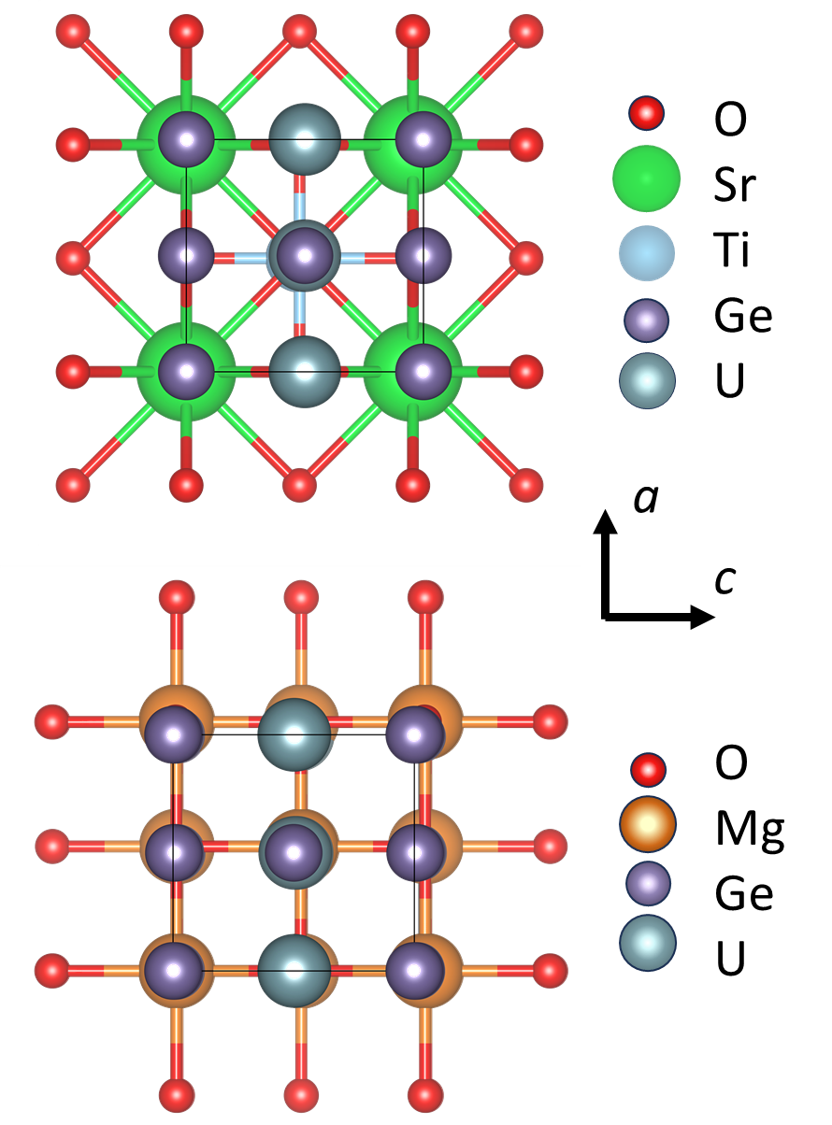}
	\caption{Possible epitaxial matches between UGe$_2$ on the $\{100\}$ surfaces of SrTiO$_3$ (STO, top) and MgO (bottom). Lattice directions $a$ and $c$ are shown for UGe$_2$. Images created using VESTA \cite{vesta}.}
	\label{match_figure}
\end{figure}

 Focusing on the UGe$_2$ system, although the system is formally orthorhombic, the $a$ and $c$ lattice parameters differ only by $\sim$ 2 \%, giving the possibility of a relatively simple `cube-on-cube' epitaxial relationship with common cubic substrates. Two examples are shown in Fig$.$ \ref{match_figure}, where matches with SrTiO$_3$ (STO) $\{100\}$ and MgO $\{100\}$ surfaces are schematically shown. In the case of STO, (lattice parameter $a = 3.906$ \AA) there would be a $-2.3$\% strain (compressive) for the $a$ direction of UGe$_2$, and a $+4.7$\% strain (tensile) in the $c$ direction, assuming coherent growth. Here the mismatch has been defined as $100$\% $\times (a_{\mathrm{film}} - a_{\mathrm{substrate}}) / a_{\mathrm{substrate}}$. For MgO (lattice parameter $a = 4.213$ \AA) the strains are $\sim +3$\% and $+5$\% (both tensile) for $c$ and $a$, respectively. The lattice mismatch between cubic UGe$_3$ and these two substrates assuming a cube-on-cube epitaxial relationship is also relatively small for the MgO in particular ($+0.17$\%), and $-7.7$\% for STO. These possible matches motivated this study. Finally we note the $\{100\}$ surfaces of CaF$_2$ (lattice parameter $a = 5.462$ \AA).  While not as well-matched to the various U-Ge phases for simple epitaxial relationships, we also used CaF$_2$ in this work as a comparison, since it lacks oxygen in its structure. As will be subsequently shown in this work, the oxygen can be gettered by the U-contining film from the substrates during growth at elevated temperatures, leading to unwanted oxide phases. 

\section{Experimental Methods}
\label{methods}
All samples in this study were grown using the dedicated `FaRMS' actinide d$.$c$.$ magnetron sputtering capability at the University of Bristol, UK \cite{farms_website}. This ultra-high vacuum system operates at base pressures $\sim 10^{-10}$ mbar and contains four sputtering guns inside a load-locked chamber. Substrates are loaded onto an adjustable height stage adjacent to a Ta resistive heater capable of achieving temperatures of up to $850$ $^\circ$C. Samples were fabricated by co-sputtering from separate U and Ge targets in presence of Ar at a pressure of $7.3\times10^{-3}$ mbar. Deposition rates for U and Ge were individually calibrated by ex-situ x-ray reflectivity measurements. Typical deposition rates were $0.03 - 0.06$ nm/s, and were tuned by varying the power suppled to each target in the range $12-30$ W for Ge, with the U gun power held at $10$ W. Nominally for a 1:2 U:Ge ratio the estimated required power ratio was also 1:2. For samples grown above room temperature (RT), the substrates were held at the growth temperature, $T_{\mathrm{g}}$, to within an error of $\pm 25$ $^{\circ}$C. Film thicknesses were nominally $\sim$ 50 nm for all samples. A polycrystalline Nb cap, of thickness $\sim 9$ nm was grown in-situ on all samples after cooling the samples to ambient temperature, to prevent oxidation in air. 

Three types of commercial cubic single crystal substrate were used for growth: CaF$_2$, SrTiO$_3$, and MgO. All substrates were purchased from Crystal GmbH, had dimensions $10$ mm $\times$ $10$ mm $\times$ $0.5$ mm, were oriented with a $<100>$ direction normal to the substrate plane, and were polished to optical grade. Table \ref{UGe_deposition_conditions} summarises the samples fabricated in this study. X-ray diffraction (XRD) experiments were carried out at room temperature with a laboratory based Cu-K$_{\alpha}$ source to determine the U-Ge phases present in the samples, as well as their crystallographic orientation in the specular direction. Temperature-dependent transport measurements were carried out in a liquid helium-4 dip probe, in a four-point resistance geometry, using a d.c. bipolar bias current method with current magnitude of the order of 100 $\mu$A. 
\begin{table}[ht]
    \centering
    \begin{tabular}{|c|c|c|c|c|}
        \toprule
        Substrate & Sample ID & \multicolumn{2}{|l|}{Power (W)}   & $T_{\mathrm{g}}$ ($^\circ$C) \\
          &  & U & Ge  &   \\
                \midrule
        STO  & SN-1 & 10 & 20 & RT \\
            & SN-2 & 10 & 20 & $300$  \\
            & SN-3 & 10 & 20 & $525$  \\
        \midrule
        MgO  & SN-4 & 10 & 20 & RT \\
            & SN-5 & 10 & 20 & $300$  \\
            & SN-6 & 10 & 20 & $525$    \\
            & SN-7 & 10 & 20 & $775$  \\
       & SN-8 & 10 & 20 & $850$ \\
            & SN-9 & 10 & 15 & $850$ \\
            & SN-10 & 10 & 12 & $850$ \\
            \midrule
        CaF$_2$  & SN-11 & 10 & 30 & $525$  \\
            & SN-12 & 10 & 25 & $525$  \\
            & SN-13 & 10 & 20 & $525$ \\
            & SN-14 & 10 & 17 & $525$   \\
        \bottomrule
    \end{tabular}
    \caption{Summary of samples grown on various substrates with varying growth temperatures, $T_{\mathrm{g}}$, and sputtering powers.}
    \label{UGe_deposition_conditions}
\end{table}

\section{Results \& Discussion}
\label{results}

\subsection{Temperature series on SrTiO$_3$}
\label{sec:7.3}
\begin{figure}[h]
    \centering
    \includegraphics[width=0.8\linewidth]{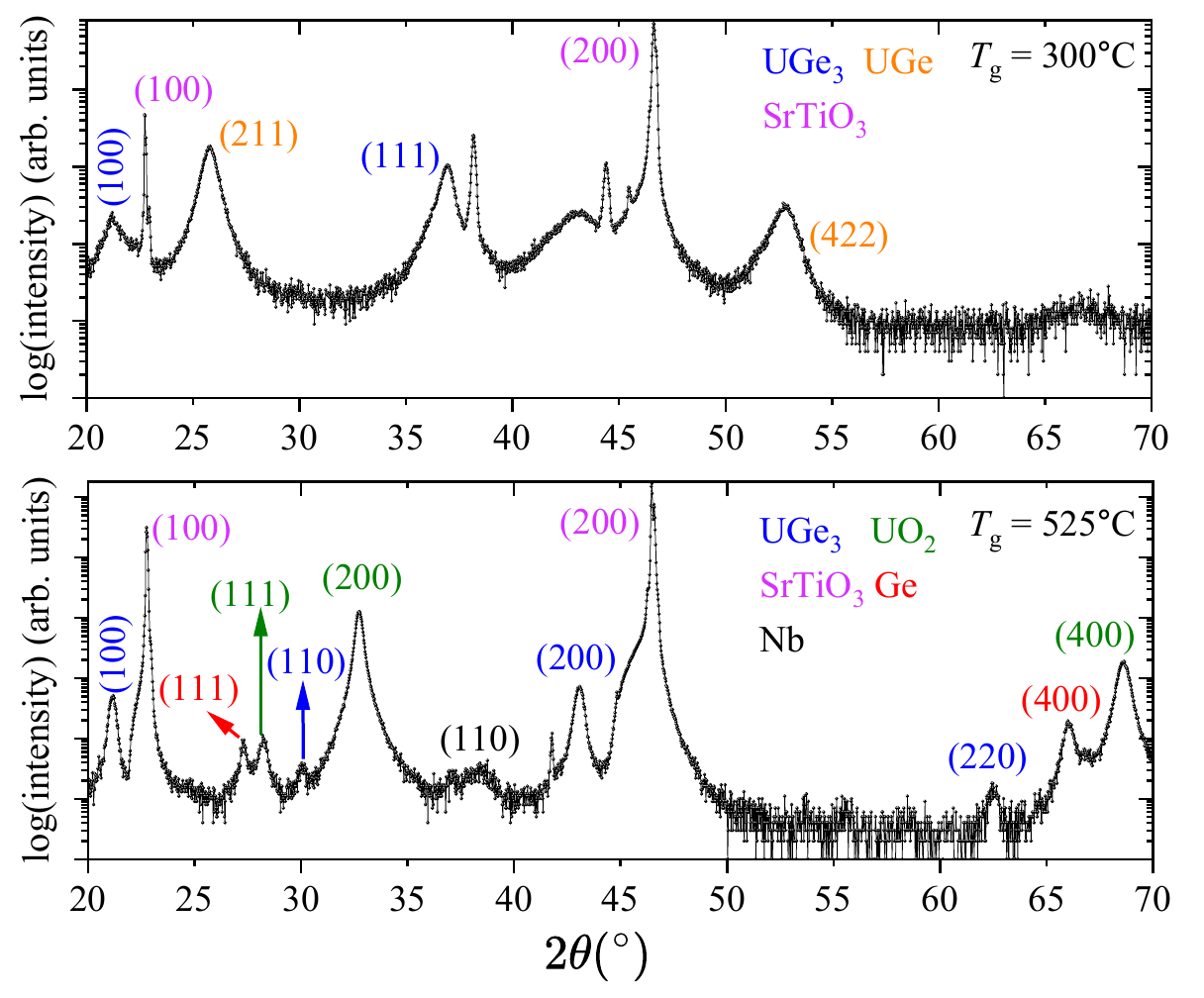}
    \caption{Specular XRD measurement of U-Ge thin films grown on STO for $T_{\mathrm{g}} = 300$ $^\circ$C (top) and $525$ $^\circ$C (bottom). Indexed planes are colour-coded according to the corresponding material. Lines are a guide to the eye.}
    \label{STO temperature study}
\end{figure}

Samples were grown at three values of $T_{\mathrm{g}}$: RT, $300$ $^{\circ}$C and $525$ $^{\circ}$C, at constant sputtering power onto STO $\{100\}$ substrates. The sample grown at RT showed no significant peaks in the XRD measurement other than those associated with the substrate and the data are not shown. $2\theta-\omega$ XRD scans for the two samples grown at higher $T_{\mathrm{g}}$ are shown in Fig$.$ \ref{STO temperature study}.
For the sample grown at $T_{\mathrm{g}} = 300$ $^\circ$C reflections from two U-Ge phases (UGe$_3$ and UGe) were identified, with two dominant different orientations present for UGe$_3$, and one for UGe. When $T_{\mathrm{g}}$ was raised to $525$ $^\circ$C, several changes to the XRD data are clear. Firstly UO$_2$ peaks start to appear particularly associated with reflections from the (200) plane with some lower intensity peaks related to reflections from (111) planes. The UGe phase is no longer present, and additionally elemental Ge and Nb reflections are observed. 

Several of these observations are linked. STO is well known to be sensitive to loss of oxygen at elevated temperatures and in vacuum (e.g. \cite{spinelli_prb_2010}), and we therefore conclude that the oxygen required to form UO$_2$ originates from the STO substrate. This in turn reduces the amount of readily available U-atoms to form U-Ge compounds, reducing the UGe component. With the loss of the U-atoms to form U-Ge phases, the excess Ge atoms crystallises in its elemental form. It should also be highlighted that although both samples contained UGe$_3$ phases, the higher temperature sample showed that the phase is $\{100\}$ textured. Finally we note that the Nb $(110)$ peak is visible in the higher temperature sample only since the $(111)$ UGe$_3$ Bragg peak obscures the signal in the lower $T_{\mathrm{g}}$ sample. 

\begin{figure}[]
    \centering
    \includegraphics[width=0.8\linewidth]{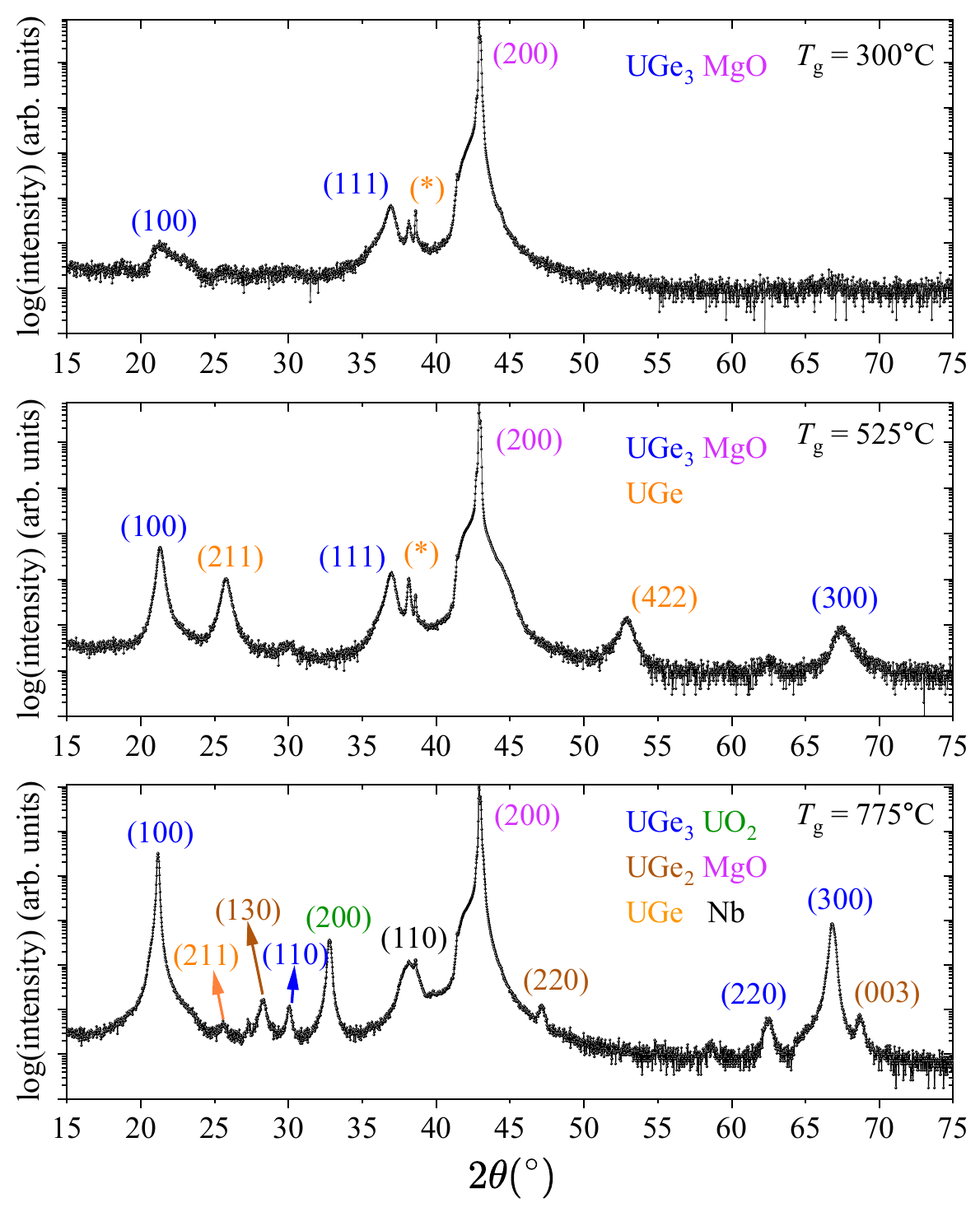}
    \caption{High angle XRD measurement of U:Ge thin films grown on MgO with varying temperatures. Indexed planes are colour-coded according to the corresponding material. At higher temperatures, as in the case of STO, UO$_2$ peaks start to appear. Stars indicate peaks that could not be indexed to any expected phases.}
    \label{MgO temperature study}
\end{figure}

\subsection{Temperature and power series on MgO}
\label{highert}
Motivated by the epitaxial USi$_2$ film grown on MgO substrates, and the potential epitaxial match discussed above, five U-Ge samples were grown at temperatures varying between RT up to $T_{\mathrm{g}}=850$ $^{\circ}$C. The same U:Ge sputtering power ratio was used as for the STO samples, as well as an additional two samples at $T_{\mathrm{g}}=850$ $^{\circ}$C with varying Ge sputtering power.

First we focus on the four lower $T_{\mathrm{g}}$ samples. The $2\theta-\omega$ XRD scans for these samples are shown in Fig$.$ \ref{MgO temperature study}. As for the samples grown on STO, the RT grown sample showed no Bragg peaks (data not shown). Also at the highest of these four growth temperatures, $T_{\mathrm{g}}= 775^{\circ}$C, Bragg peaks corresponding to UO$_2$ $(200)$ were observed. Again both UGe and UGe$_3$ phases were observed. The UGe$_3$ phase showed $(100)$ texture when $T_{\mathrm{g}}$ was increased to $775^{\circ}$C, with the $(111)$ orientation observed at lower $T_{\mathrm{g}}$. However in the case of the samples grown on MgO substrates there was no evidence of crystalline elemental Ge, which was observed in the STO series. 

A relatively weak signal from the UGe$_2$ phase can also be observed for the sample grown on MgO at $\sim 775$ $^{\circ}$C. Increasing $T_{\mathrm{g}}$ from $\sim 525$ $^{\circ}$C to $775$ $^{\circ}$C suppresses the UGe phase, similar to what was observed using the STO substrate. For the sample grown at $775$ $^{\circ}$C assuming a cubic lattice for UGe$_3$, using the specular $(100)$ and $(300)$ Bragg peaks at $21.17^{\circ}$ and $66.85^{\circ}$ respectively, and $d_{hkl} = a/\sqrt{h^2+ k^2 + l^2}$, the lattice parameter can be estimated to be $a=4.197$ \AA, close to the bulk value. 
\begin{figure}[h]
    \centering
    \includegraphics[width=0.8\linewidth]{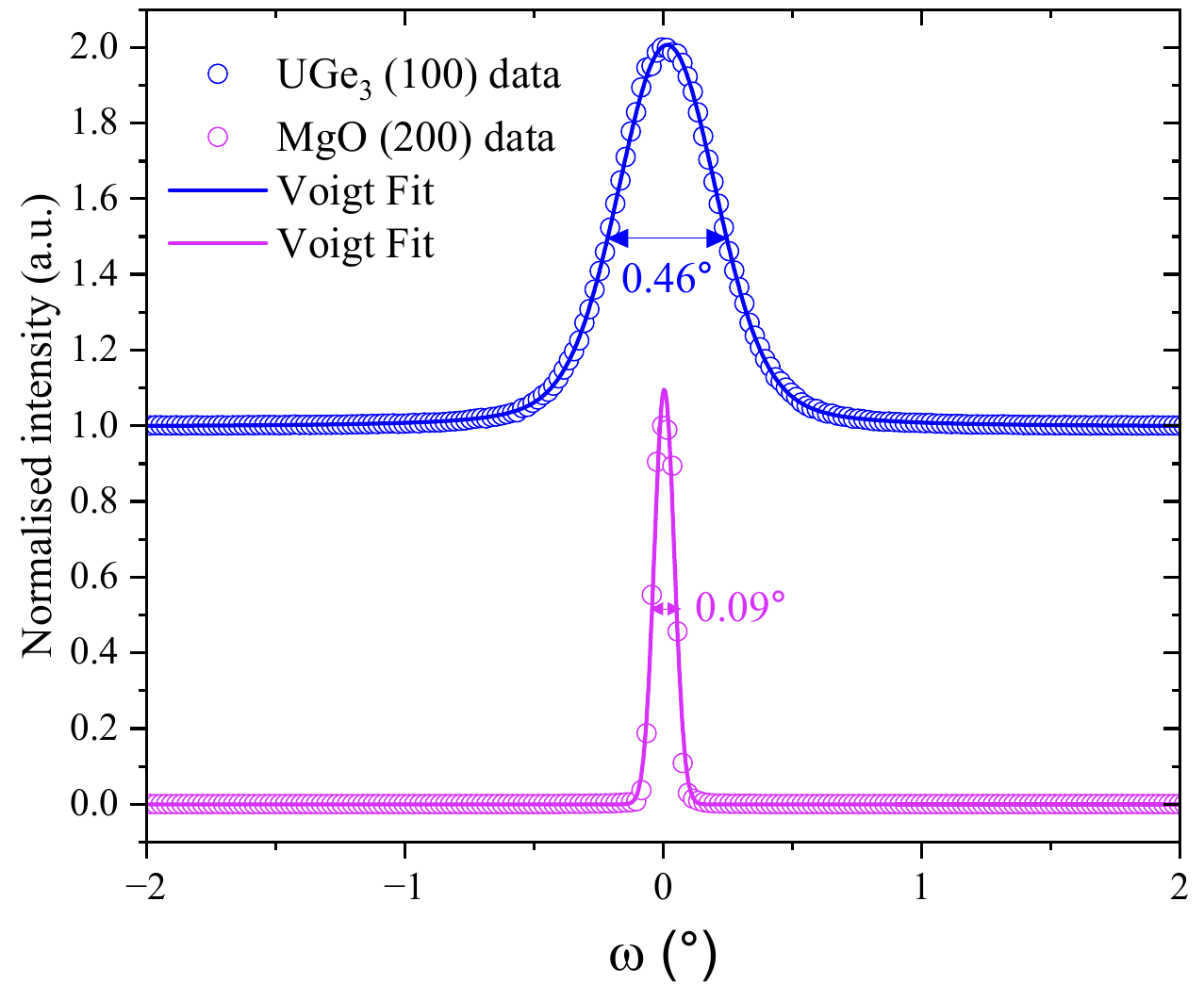}
    \caption{$\omega$ scan comparison of UGe$_3$ $(100)$ and MgO $(200)$ peaks for sample SN-7. Open circles are the data and the solid lines are best fits using the Voigt function. UGe$_3$ data are offset vertically for clarity.}
    \label{RC on MgO and UGe3}
\end{figure}

\begin{figure}[h]
			\centering
		\includegraphics[width=0.8\linewidth]{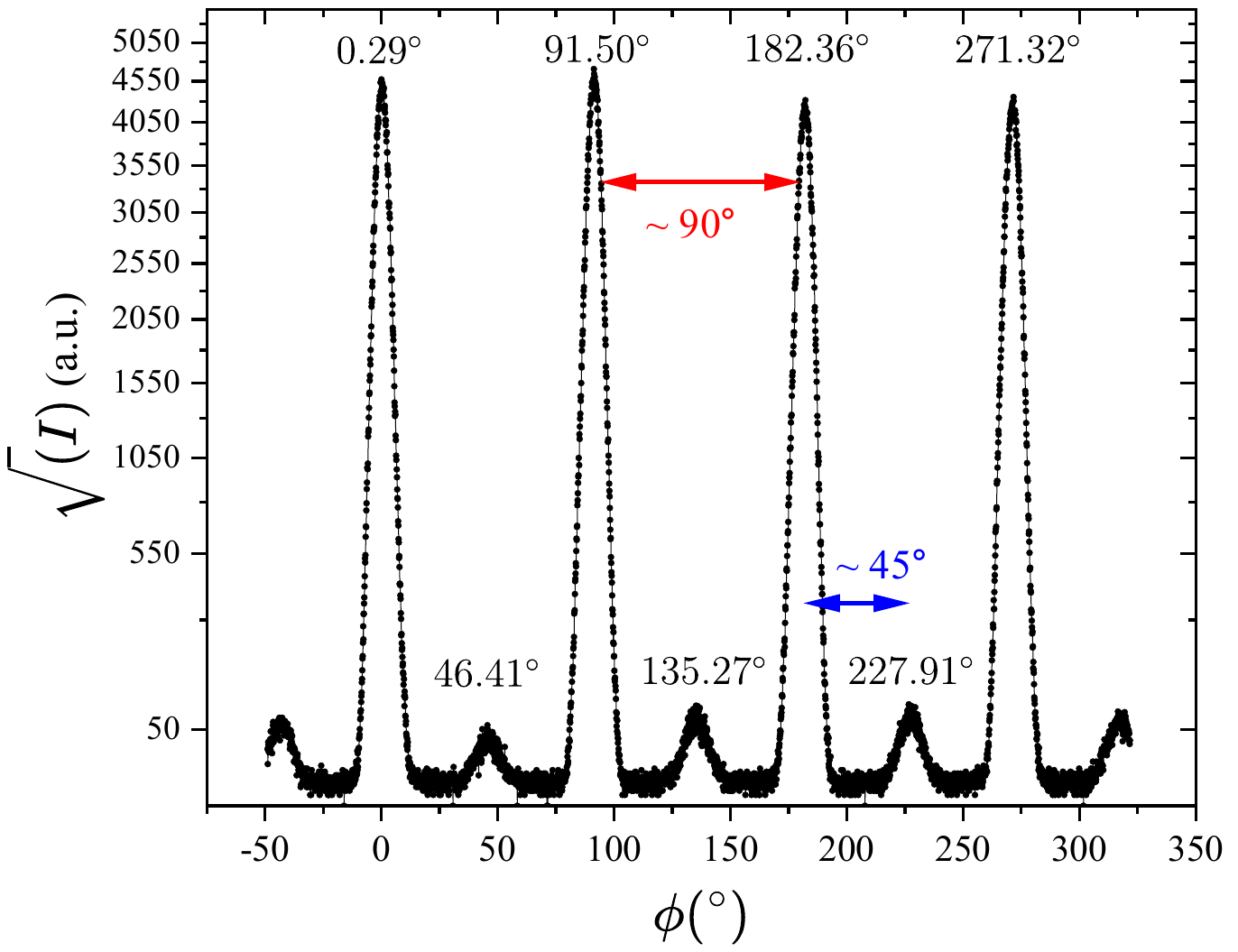}
		\caption{Pole figure data for the \{310\} reflections of UGe$_3$ for sample SN-7.}
		\label{UGe3 Off specular scan for MgO}
	\end{figure}

In addition to the coupled scan presented, an $\omega$ scan provides information on the crystal mosaicity or tilt disorder, as well as the presence of strain, defects, or asymmetric domain distributions in the system. For the sample grown at $775$ $^{\circ}$C on MgO, the full width at half maximum (FWHM) of the (100) UGe$_3$ peak was $0.46\pm0.01^{\circ}$, (the FWHM of the (200) MgO substrate peak was $0.091\pm0.001^{\circ}$), indicating good crystallinity of the film. For this sample, off-specular XRD scans were conducted to investigate the in-plane matching with the substrate, as shown in Fig.~\ref{UGe3 Off specular scan for MgO} for the $\{310\}$ reflection family of UGe$_3$. For a bulk single crystal of UGe$_3$ with the {\it fcc} structure, one would expect to see a $4-$fold symmetry in a $360^{\circ}$ rotation in $\phi$. These peaks should be symmetric and have a periodicity of $\Delta\phi=90^{\circ}$ between each peak. In fact the off-specular scan shows eight distinct peaks alternating in $\phi$ between relatively high and low intensity. These two sets of peaks indicate the presence of a major domain and a minor domain of UGe$_3$ rotated $45^{\circ}$ with respect to each other. Given the discussion at the beginning of this paper on the possible good epitaxial match with this substrate, the formation of UGe$_3$ on MgO with a cube-on-cube match was predicted, but the minor domains rotated by $45^{\circ}$ were not expected.   

\begin{figure}[]
    \centering
    \includegraphics[width=0.8\linewidth]{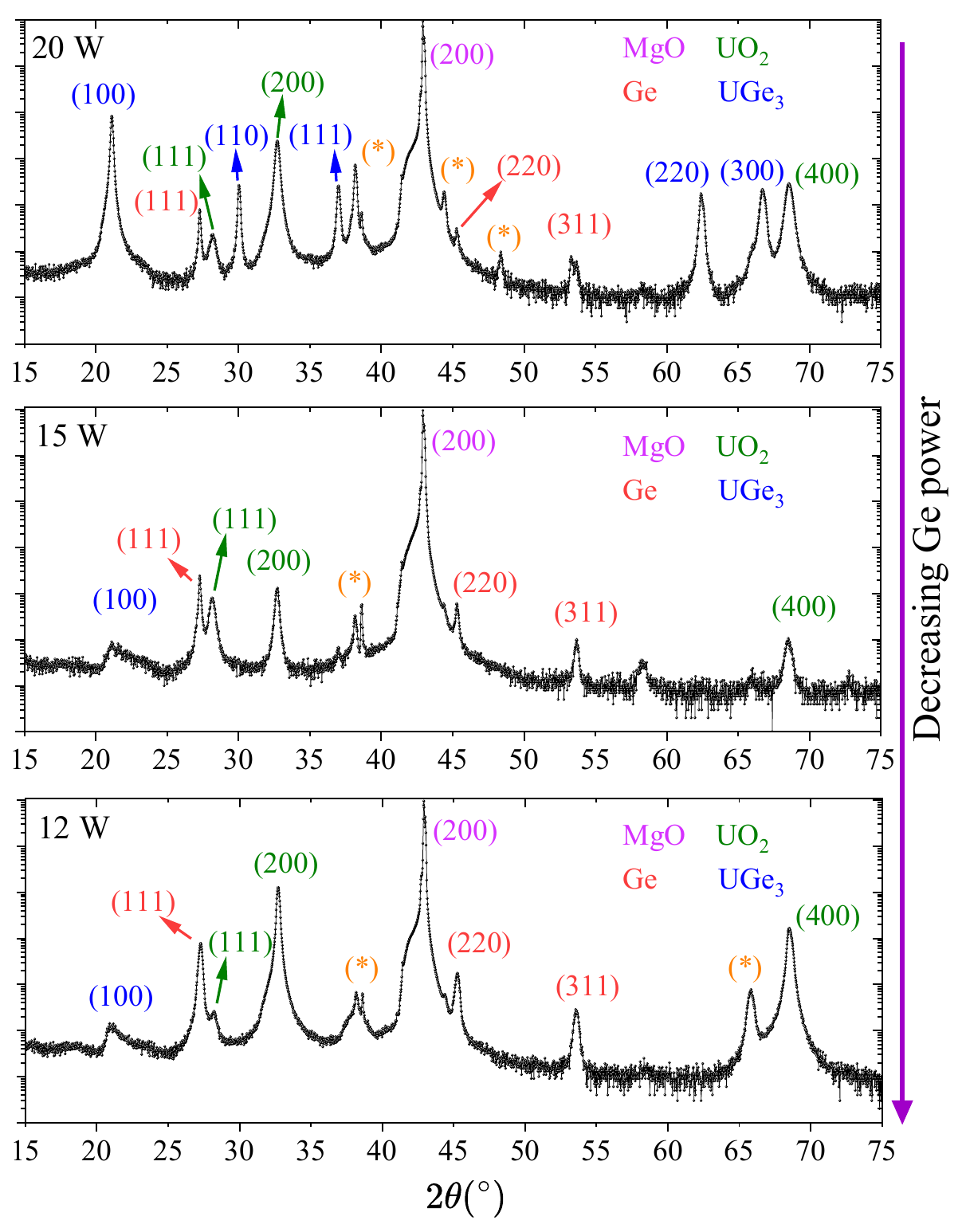}
    \caption{High angle measurement of U:Ge thin films grown on MgO at $\sim850$ $^{\circ}$C with varying Ge sputtering power indicated in the top left of each panel. Indexed planes are colour-coded according to the corresponding material. At lower Ge power, relatively stronger elemental Ge peaks become visible while UGe$_3$ peaks are suppressed. Stars indicate peaks that could not be indexed to any expected phases.}
    \label{MgO high temp power series study}
\end{figure}

To further study the U:Ge thin phases on the MgO substrates, a power series study was conducted at a slightly higher temperature of $850$ $^{\circ}$C, with variable Ge sputtering power and fixed U sputtering power (see the growth details in Table \ref{UGe_deposition_conditions}). The specular XRD scans of these samples are shown in Fig$.$ \ref{MgO high temp power series study}. 
In all three XRD scans there is clear evidence of the UGe$_3$ phase, with several different growth orientations for the highest Ge power, but with UGe$_3$ $(100)$ reflections present for all three samples. The formation of UO$_2$ is also clear in the XRD data with $(200)$, $(400)$ and $(111)$ peaks all observable, consistent with the observations for the $T_{\mathrm{g}} = 775$ $^{\circ}$C sample, SN-7. Similar to sample SN-3 grown on STO (although at a higher $T_{\mathrm{g}}$), here we again find peaks corresponding to elemental Ge in the specular scans in all three samples. It is also notable that the reduction in Ge sputtering power does not drive the formation of UGe phase. As we will see in the next section such a trend does occur when CaF$_2$ substrates were used.  

\subsection{CaF$_2$ substrate}
\begin{figure}[h]
    \centering
    \includegraphics[width=0.8\linewidth]{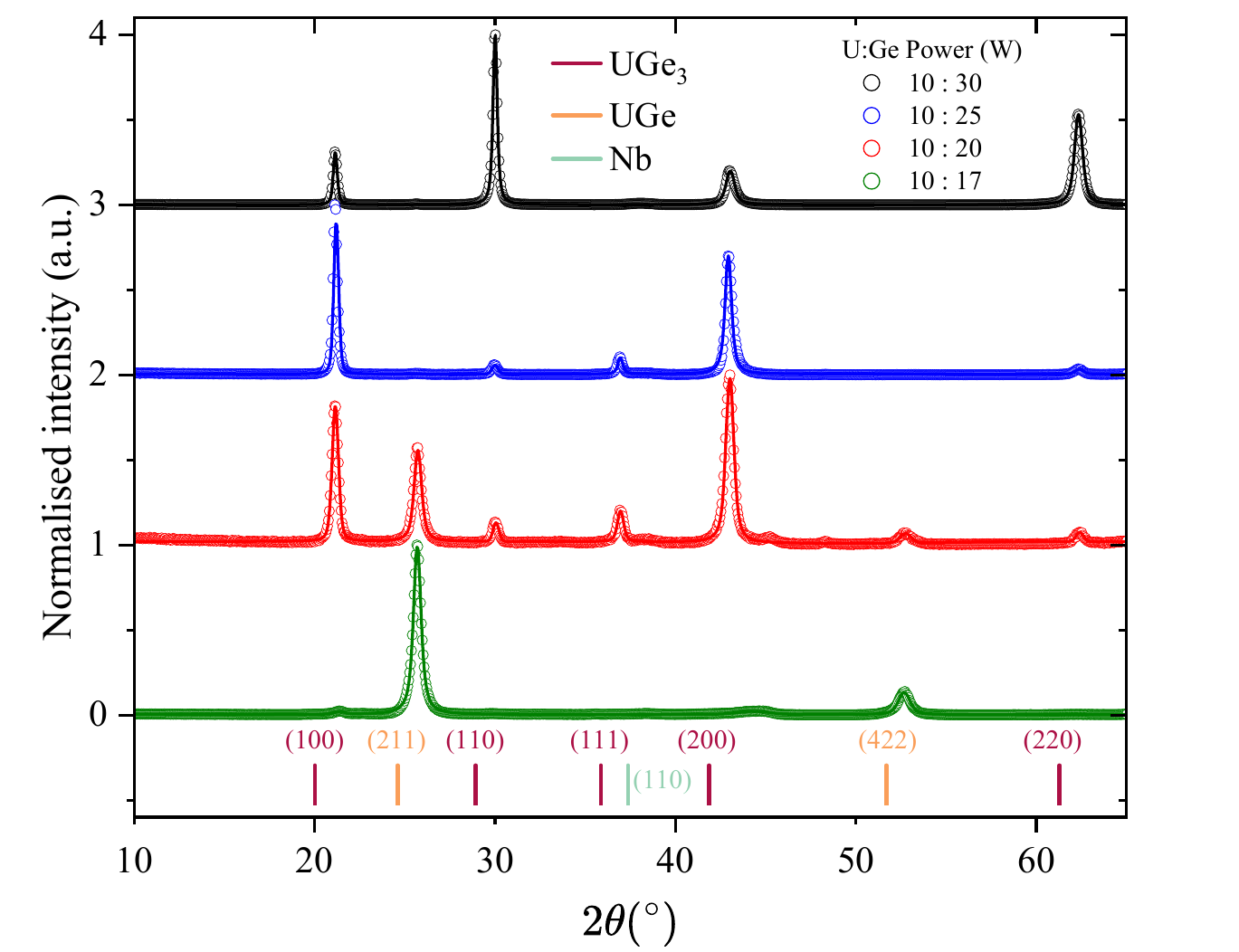}
    \caption{Specular XRD measurements of U-Ge thin films grown on CaF$_2$ with varying Ge growth power. The data have been normalised and offset vertically for clarity. Ge content decreases from top to bottom.  Colour-coded vertical lines at the bottom show the literature peak positions for the various reflections for the different phases present in the films.}
    \label{CaF2 power series study}
\end{figure}

Finally in this section we discuss the XRD characterisation of the samples grown on $(100)$ CaF$_2$ substrates, which were deposited at the same $T_{\mathrm{g}} = 525$ $^{\circ}$C with a range of Ge sputtering powers. The normalised XRD data for these samples are shown in Fig$.$ \ref{CaF2 power series study}. Note that the $(200)$ CaF$_2$ substrate reflection at $\sim32.76^{\circ}$ cannot be seen here since an $\omega$ angle offset was introduced to reduce the substrate peak intensity, making it easier to analyse the other features in the dataset. 
For the sample with the highest Ge content (SN-11), the major Bragg peaks observed are indexed to UGe$_3$ with $(110)$ texture. On the other hand, sample SN-14, grown with the lowest Ge power shows stronger Bragg peaks from the $(211)$ plane of UGe. This is consistent with the reduction in Ge content in these thin films moving across the phase diagram. In the case of samples SN-12 and SN-13, grown with intermediate Ge power, both UGe and UGe$_3$ phases are present. On closer inspection, there is also a texture change: the UGe$_3$ phases show $(100)$ texture in these two samples instead of $(110)$ seen in the Ge-richer sample. Additionally, it can also be seen that between the two mixed-phase samples, the Ge-poorer sample (SN-13), indicates the transition point where the (211) UGe peaks clearly starts to appear alongside the (100) UGe$_3$ peak seen. No other U-Ge phases were observed in these samples.
\begin{figure}[]
    \centering
    \includegraphics[width=0.8\linewidth]{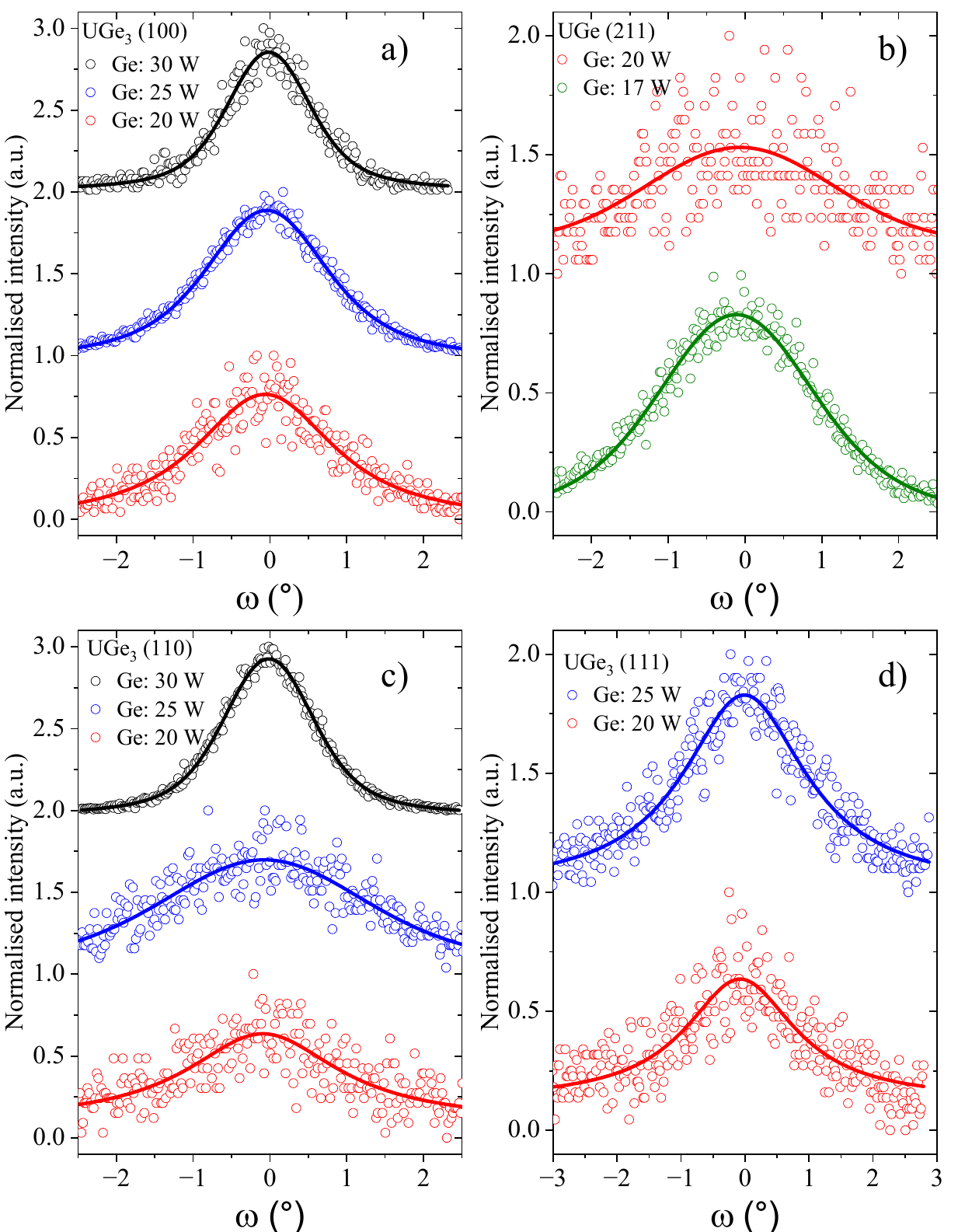}
    \caption{$\omega$ scan on selected UGe$_3$ and UGe peaks for samples grown on CaF$_2$ with various Ge powers. \textbf{a)} (100) UGe$_3$, \textbf{b)} (211) UGe, \textbf{c)} (110) and \textbf{d)} (111) UGe$_3$ peaks. The plots have been arranged in order of decreasing Ge power with the same colour code representing the samples as in the coupled scan shown in Fig. \ref{CaF2 power series study}.}
    \label{RC of UGe and UGe3 on CaF}
\end{figure}

\begin{table*}
 \centering
 \begin{tabular}{|c|c|c|c|c|c|} 
 \toprule
       Sample ID & Ge Power (W)  & \multicolumn{4}{|c|}{FWHM ($^{\circ}$)} \\
         &   & (100) UGe$_3$  & (211) UGe & (110) UGe$_3$  & (111) UGe$_3$ \\
       \midrule
        SN-11 & $30$ & $1.32\pm0.03^{\circ}$ & $-$& $1.43\pm0.01^{\circ}$ & $-$ \\
        SN-12 & $25$ & $1.94\pm0.03^{\circ}$ & $-$ & $3\pm1^{\circ}$ & $2.3\pm0.9^{\circ}$ \\
        SN-13 & $20$ & $2.2\pm0.1^{\circ}$ & $2.83\pm0.3^{\circ}$  & $2.4\pm0.1^{\circ}$ & $2.1\pm0.5^{\circ}$ \\
        SN-14 & $17$ & $-$ & $2.52\pm0.07^{\circ}$ & $-$ & $-$\\
        \bottomrule
  \end{tabular}
  \caption{FWHM values for selected peaks of UGe$_3$ and UGe thin films grown on CaF$_2$ calculated based on the Voigt functional fits made to the $\omega$ scans of $(100)$, $(110)$ and $(111)$ UGe$_3$ peaks and the $(211)$ UGe peak.}
\label{FWHM of UGe peaks on CaF2}
\end{table*}

To quantify the mosaicity and crystallinity of the samples, $\omega$ XRD scans were made on selected UGe$_3$ and UGe reflections. These $\omega$ scans are shown in Fig. \ref{RC of UGe and UGe3 on CaF}. After fitting Voigt functional forms to the data, the resultant FWHM of the peaks are summarised in Table \ref{FWHM of UGe peaks on CaF2}. Firstly, focusing on the $(100)$ UGe$_3$ peak, we can see that the FWHM decreases with decreasing Ge power indicating a reduction in crystallinity. For the $(110)$ UGe$_3$ peak the $30$ W sample (SN-11) has the best crystallinity of the series. Interestingly for this sample, while we see an improvement in the texture compared to the other samples, we did not observe an improved rocking curve. When focusing on the $(211)$ UGe peak, the coupled scan FWHM clearly improves for the lower Ge power (i.e. moving from the sample grown at $20$ W to the sample grown at $17$ W). This change is also reflected in the $\omega$ scan: a slightly narrower FWHM peak. Finally we note that the $(111)$ reflection of UGe$_3$, when visible (only for the two intermediate growth powers) reveal little difference in crystallinity to within the measurement error.  

\section{Resistivity measurements}
\label{resistivity}

\begin{figure*}[!h]
    \centering
    \includegraphics[width=\linewidth]{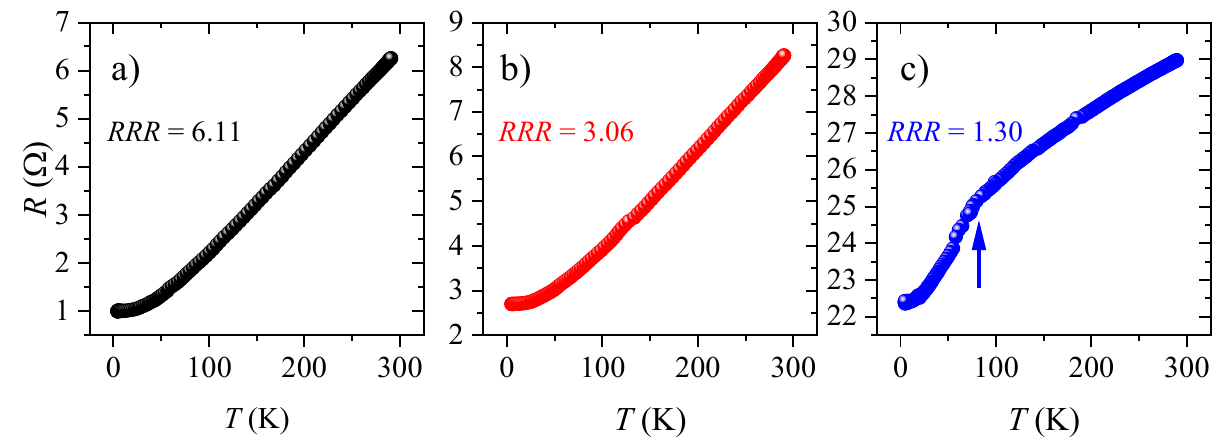}
    \caption{Resistance-temperature measurements on selected UGe thin films.\textbf{a)} Sample SN-7 grown on MgO [001], which showed majority UGe$_3$ peaks. \textbf{b)} SN-11 grown on CaF$_2$ [001], which showed majority UGe$_3$ peaks. \textbf{c)} Sample SN-14, grown on CaF$_2$ [001], which showed strong UGe peaks. Blue arrow in (c) indicates a distinct kink in the $R(T)$ data. Residual resistance ratios ($RRR$) are shown in each panel.}
    \label{UGeRT}
\end{figure*}
In addition to the XRD measurements presented, some longitudninal resistance versus temperature ($R(T)$) measurements were performed on three selected samples as shown in \ref{UGeRT}. These measurements were performed a He-4 dip probe, in a temperature range of $4.2$ K $\leq T\leq290$ K. Firstly all the samples showed metallic behaviour with a rather linear $R(T)$ at higher temperatures. As a measure of the quality of the samples, we show the residual resistance ratio ($RRR$), given by $RRR = R(T = 295 \mathrm{K}/ R(T = 4.2 \mathrm{K})$. Comparing the two UGe$_3$ samples, the sample grown on the MgO substrate showed a larger $RRR \sim 6.1$ in comparison to the sample grown on CaF$_2$ with $RRR \sim 3.1$, consistent with the narrower rocking curve of the UGe$_3$ peaks for the sample grown on the MgO substrate. Note that bulk UGe$_3$ does not show any particular magnetic ordering down to 2 K, so the feature-less $R(T)$ plots are consistent with this. On the other hand the $R(T)$ for the sample with the majority UGe phase shows a distinct change in the slope of the $R(T)$ at around $\sim74$ K which possibly indicates the onset of some spin fluctuations, consistent with previous specific heat measurements that have indicated a larger specific heat coefficient $\gamma(0)$ value for UGe in comparison to the other germanides \cite{pikul_low-temperature_2014}. Overall however, it is clear that the MgO substrate gives rise to the highest quality UGe$_3$ films, and further thicknesses are required to understand if surface scattering is limiting the value of $RRR$, or whether scattering due to the phase impurities is more dominant. 

\section{Summary and conclusions}
In conclusion we have stabilised a range of crystalline binary U-Ge phases on various substrates using thin film growth at a range of growth temperatures and sputtering powers. Although we have been unable to stabilise single phase material so far, we have observed dominant growth of UGe$_3$ and UGe phases by varying the substrate material and growth conditions. Majority UGe$_2$ phase thin films have so far proved illusive. A notable challenge was the emergence of oxide phases at higher growth temperatures due to gettering by the U of oxygen in the substrate material. Future studies should examine the possibility of buffer layers between the substrate and the U-Ge layers that can simultaneously act as a diffusion to oxygen migration, as well as providing further degrees of freedom for epitaxial matching and phase stabilisation. We note after the completion of this work, related spectroscopic studies of U-Ge samples were published elsewhere \cite{alexa_uge2_2026}. 
\label{conclusions}

\section*{Acknowledgements}
We acknowledge the funding and support from the UKRI Engineering and Physical Sciences Research Council (EPSRC), UK. In particular this work utilised the National Nuclear User Facility FaRMS, grant: EP/V035495/1. We thank P. Babu and Z. Huang for early help in this project, and R. Nicholls for insightful comments. \\

\bibliographystyle{unsrt} 
\bibliography{example}
\end{document}